# Trends in European flood risk over the past 150 years

Dominik Paprotny[1,2], Antonia Sebastian[1,3], Oswaldo Morales-Nápoles[1], Sebastiaan N. Jonkman[1]

1. Department of Hydraulic Engineering, Faculty of Civil Engineering and Geosciences, Delft University of Technology, Stevinweg 1, 2628 CN Delft, The Netherlands
2. European Commission, Joint Research Centre (JRC), Directorate E – Space, Security & Migration, Disaster Risk Management, via E. Fermi 2749, I-21027 Ispra (VA), Italy.
3. Department of Civil and Environmental Engineering, Rice University, 6100 Main Street, Houston, Texas 77005, USA.

Correspondence: Dominik Paprotny (d.paprotny@tudelft.nl)

Flood risk changes in time and is influenced by both natural and socio-economic trends and interactions. In Europe, previous studies of historical flood losses corrected for demographic and economic growth ('normalized') have been limited in temporal and spatial extent, leading to an incomplete representation in trends of losses over time. In this study we utilize a gridded reconstruction of flood exposure in 37 European countries and a new database of damaging floods since 1870. Our results indicate that since 1870 there has been an increase in annually inundated area and number of persons affected, contrasted by a substantial decrease in flood fatalities, after correcting for change in flood exposure. For more recent decades we also found a considerable decline in financial losses per year. We estimate, however, that there is large underreporting of smaller floods beyond most recent years, and show that underreporting has a substantial impact on observed trends.

## Contents

## Introduction

Extreme hydrological events are generally predicated to become more frequent and damaging in Europe due to warming climate[1 2 3 4 5 6 7]. Though the trajectory of future climatic developments seem certain, there is less confidence in the changes in flood risk as a result of climate change so far [8 9 10 11]. Qualitative and quantitative hydrological studies for Europe have indicated no general continental-wide trend in river flood occurrences, extreme precipitation, or annual maxima of runoff[12 13 14]. However, substantial variations between different catchments have been observed, ranging from an increase in north-western Europe to no trend or a decrease in other parts of the continent[15 16]. Similar findings were reported for storminess along the European coasts[17 18].

Natural hazards are phenomena that inherently involve adverse consequences to society. Therefore, analyses of long-term trends in flood risk should also account for changes in size and distribution of population and assets[19 20]. Without correcting reported losses for spatial and temporal changes in exposure, a significant upward trend in losses is indicated[21,22,23]. However, after adjusting nominal losses by demographic and economic growth, no significant trends for floods, both on European scale[24,25] and for individual countries were observed[26,27,28]. Such 'normalization' processes have also proven to be important for explaining trends in other natural hazards[29,30,31].

Still, there are two main limitations in existing analyses. First, historical disaster loss data are not temporally homogenous, with the number of events for which quantitative information is available declining quickly when moving back in time. The starting point for many studies is in the vicinity of the year 1970 or later. International databases of natural hazards (EM-DAT[32], NatCatService[33], Dartmouth Flood Observatory[34] or European Environment Agency[35]) provide reasonable coverage only beginning with the 1980s. Comprehensive and publicly-available national repositories of disaster loss data are few in Europe and, those that are available, focus on flood and landslide events[36,37,38,39,40]. Moreover, the completeness and extent of information contained in existing data sets varies to a significant degree. In effect, large-scale studies usually rely entirely on the contents of global or continental databases, while national studies are shaped by the specifics of locally-available data. This leads to considerable uncertainties when examining trends at the continental scale or comparing trends between countries.

Second, in virtually all studies, socio-economic variables are considered at the national level; only Munich Re utilized a coarse 1°x1° grid of exposure data (approx. 5000–9000 km$^2$ over Europe). High resolution is of particular importance for analysing flood exposure, which is relatively limited in space: at present time less than 10% of European territory is at risk of river or coastal flooding[41]. A few national studies that analysed changes in exposure found different trends in population or housing stock inside and outside hazard zones[42,43,44,45], which shows the importance of using a sufficient resolution of the analysis. Furthermore, trends in exposure and normalization of reported losses have been carried out with many different economic variables depending on the study, such as gross domestic product (GDP), variously-defined wealth or stock of housing.

In this study, we address the aforementioned limitations (short time series and low spatial resolution) of previous assessments of flood trends for Europe using two datasets which constitute a new publicly-available database 'Historical Analysis of Natural Hazards in Europe' (HANZE)[46,47]. The first dataset (HANZE-Exposure) contains high-resolution (100 m) maps of land cover/use, population, GDP and wealth in 37 European countries and territories from 1870 to 2020. The maps were created by estimating changes in the distribution of land cover/use and population relative to the year 2011, for which detailed gridded datasets are available (100 m resolution land cover/use[48] and 1 km resolution population map[49]). Based on previously published, relatively simple and explicit methods, demographic and economic data were disaggregated to 100 m resolution, and changes in historical land use and population were modelled utilizing a large compilation of historical statistics at the regional level (see Methods section for more details). The maps allow information on change in exposure to be extracted within any defined hazard zone, such as an area at risk of flooding.

The second dataset (HANZE-Events) includes records of 1564 damaging flood events that occurred within the same domain between 1870 and 2016, and for which several quantitative variables on losses are available: area inundated, fatalities (number of persons killed or missing, presumed dead), persons affected and monetary value of damages. The spatial extent of each event, or 'footprint', is established by intersecting the 100-year flood zone under present climate conditions[50,51] with country subdivisions known to have been affected (according to European Union's NUTS level 3 classification[52]). While the 100-year flood footprint is not an accurate representation of actual flood extent, it serves as a proxy for areas with the highest risk during historical floods. This allows us to analyse demographic and economic growth within the exposed area, as well as calculate reported losses relative to potential damages. Additionally, we use copula theory to analyse the dependence structure between four different variables: area inundated, fatalities, persons affected and financial losses. The simulated data pairs were used to fill in missing information in the database and provide a better estimate of trends in flood risk. Finally, we estimate the underreporting of smaller flood events in available sources and quantify its impact on the results.

Overall, our results using the HANZE database indicate an increase in inundated area contrasted by a consistent decline in flood fatalities, with no significant trend in the number of persons affected or financial losses since 1870. However, for the period after 1950, we observe a considerable decline in fatalities and monetary losses. Moreover, we show that the majority of quantitative information regarding historical flood losses is underreported by modern sources and that this has a profound impact on calculated trends. Our results indicate that when correcting for underreporting, the annual number of flood events and persons affected has increased much less than calculated using uncorrected series (and possibly declined since the mid-20$^{th}$ century), and that financial losses have declined over time. We foresee numerous applications of the HANZE database for further studies, including an analysis of trends for other hazards, assessing the potential impacts of climate change on historical losses, and studies of individual events and their impact on flood management.

## Results

**Trends in exposure.** Between 1870 and 2015, Europe experienced substantial growth in population (129%), urban area (more than 1000%), and wealth (more than 2000% constant prices). However, there has been large variability in patterns of development between regions. In 8% of European regions (NUTS 3), the total population in 2015 was lower than in 1870. Rural population across the continent declined, and fixed assets in agriculture barely changed in contrast with large increases in wealth in housing, industry and services sectors (Supplementary Fig. 1). Most important for this study are relative trends within and outside of flood-prone areas. Since 1870, the percentage of population, GDP and wealth exposed to the 100-year flood has decreased slightly for river floods, but increased for coastal floods (Fig. 1). When analysed at the continental scale, those trends are partly caused by the aforementioned rates of demographic and economic growth between regions (Supplementary Fig. 2). As the map in Fig. 1 shows, while exposure has declined in most countries, especially those in central and northern Europe, which means that population growth in flood-prone areas has been smaller than in areas not threatened by floods. Relative exposure has increased in several western and southern European states including France, Germany, Italy and the Netherlands. In general, changes in exposure of production (measured by GDP) and wealth are in line with trends in population, with some exceptions, e.g., in Italy and Hungary, where the percentage of wealth exposed has not changed since 1870 despite growth in the relative exposure of their national populations.

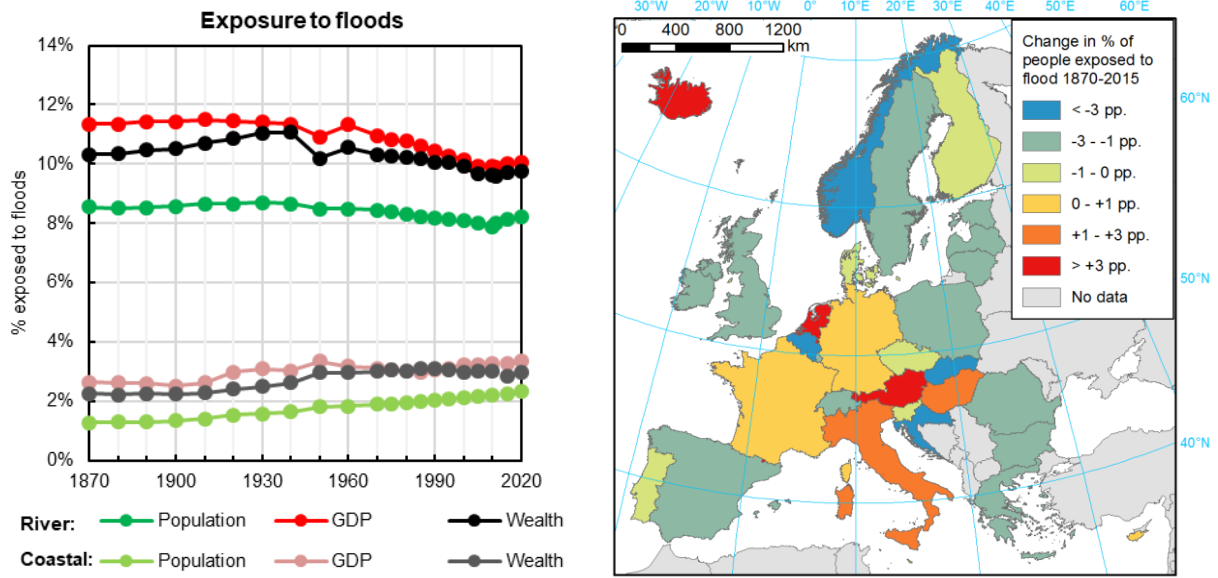

**Figure 1.** Left: Percent of the population exposed to the 100-year river and coastal flood in Europe, including short-term projection to year 2020. Right: Change in population exposed (percentage points) to the 100-year flood (either river or coastal) in each country (1870-2015).

**Distribution of flood events in Europe.** The HANZE database includes records for 1564 events (1870–2016), of which 879 (56%) are flash floods, 606 (39%) are river floods, 56 (4%) are coastal floods and the remaining 23 (1.5%) are compound events, i.e. floods caused by a co-occurrence of storm surge and high river flows. Flood events are very unevenly distributed, both during any given year and geographically (Fig. 2). In southern Europe, flash floods constituted the majority of damaging events, and were most prevalent between September and November. In central and western Europe, river floods were more frequent than flash floods, with flood losses concentrated between June and August. In northern Europe, floods were mostly caused by snowmelt and rarely resulted in significant losses. Coastal floods were mostly recorded in regions which border the North and Baltic seas.

In total, HANZE contains information on flood events that affected 1005 regions, or 74% of all NUTS3 regions within the study area. The number of floods by region is presented in Supplemental Fig. 3. On average, a flood event affected 2.8 NUTS 3 regions. The spatial distribution of floods contained in the database is heavily influenced by availability of historical records. More than half of the events in the database occurred in only three countries, namely Italy (36%), Spain (15%) and France (10%), all of which have publicly-available and searchable databases of historical floods. Thus, the large number of recorded events in those countries is a result of better coverage of events with relatively small impact on population or assets, i.e., "small floods". In contrast, total flood losses are more evenly spread out across Europe and less than a third of people affected by floods resided in the aforementioned three countries. This is partially a result of better coverage of major flood events across all countries, whereas flood events recorded in Italy, Spain and France were dominated by flash floods.

It should be noted that quantitative information on floods losses was not always obtainable. The most frequently available statistic was the number of fatalities, as they were recorded for 1547 events (99%), of which 372 events resulted in no deaths. For the remaining 17 events some fatalities were reported to have occurred, but the exact number of deaths was unknown. Information on the flooded area was only available for 157 events (10%), persons affected for 682 (44%) and monetary losses for 560 (36%).

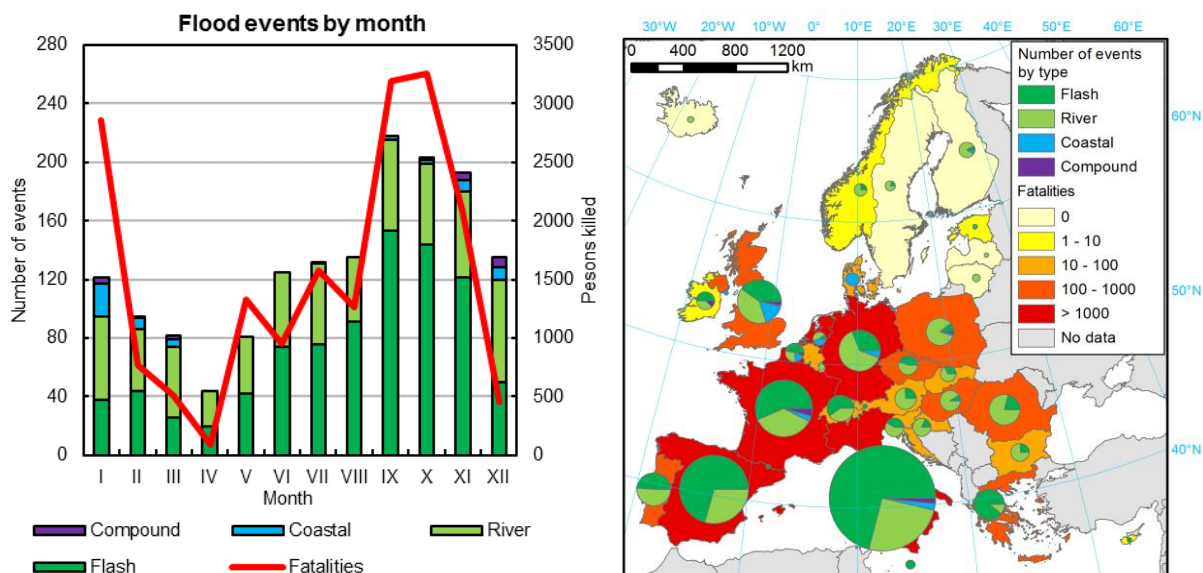

Figure 2. Total number of events and fatalities (unadjusted, reported values) between 1870 and 2016 by month (left) and by country (right).

**Trends in reported and normalized flood losses.** In Fig. 3, the records from the database are aggregated per year, and shown in two variants. In saturated colours, the original, unadjusted values of damages are shown as reported in historical records. Only the monetary value of losses was adjusted for price of inflation and converted to 2011 euros. In less intense colours, the normalized values, i.e. adjusted for change in population, GDP or wealth within the individual floods' footprints, are presented between the year of the event and 2011. It is important to note that vulnerability to floods is assumed to be constant and that the reported losses are only multiplied by the change in number of persons, production or assets in a given footprint (see Methods section for details).

The resulting trends are reported in Table 1 for five periods: 1870–2016, 1900–2016, 1930–2016, 1950–2016 and 1970–2016. Most flood events recorded in the database occurred in recent decades, with relatively small numbers of events reported for the late 19$^{th}$ century. Over most of the period of record, the total area inundated grew strongly, however no significant trend is observed during the period after 1930. Given that area flooded is known only for a tenth of all events in the database, confidence in this result is low. In contrast, the number of fatalities is available for almost all flood events in the database and a negative trend of at least 1% per year is observed, even though it is only statistically significant for the period between 1950 and 2016 (see Methods section for statistical significance testing procedure). Finally, for both the number of persons affected and monetary losses adjusted for inflation, a positive trend is observed over all periods of record. However, for 1950–2016 and 1970–2016 the trend is not significant.

Normalization has a considerable effect on the observed results. The downward trend in fatalities becomes much more pronounced, reaching -4.6% per year (1950–2016). It also becomes statistically significant except for the period between 1970–2016; however, uncertainty regarding past exposure to floods renders the trends for this time period insignificant. Nonetheless, during the period from the 1980s to the present there have been fewer (normalized) deaths than almost any prior period. In contrast, the number of persons affected increases consistently throughout time, but the trend is less pronounced than before normalization (approximately 1% per year compared to almost 2% without adjustment). Still, the number of victims peak around the year 2000. In terms of financial losses, the increase for 1870–2016 becomes smaller after normalization (1.4–1.5% per year instead of 3%), but still significant. However, when using the starting years 1900 and 1930 for the analysis, the trend in financial losses becomes statistically non-significant. The biggest shift occurs for the period between 1950 and 2016 where the trend (-2.6% per year) is statistically significant. This is similar to the situation before normalization, however the trend is now downward rather than upward. Correcting losses by changes in both GDP and wealth indicates that losses peaked in the 1950s rather than the 2000s. In general, flood losses have been declining in the entire post-1945 period despite some noticeable cycles of higher and lower loss-generating periods.

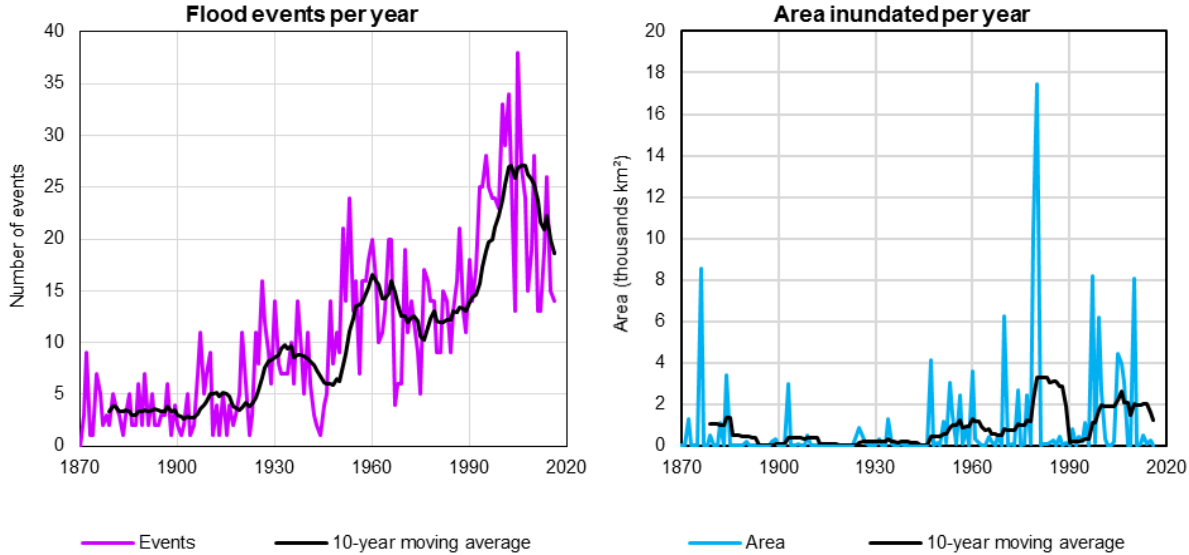

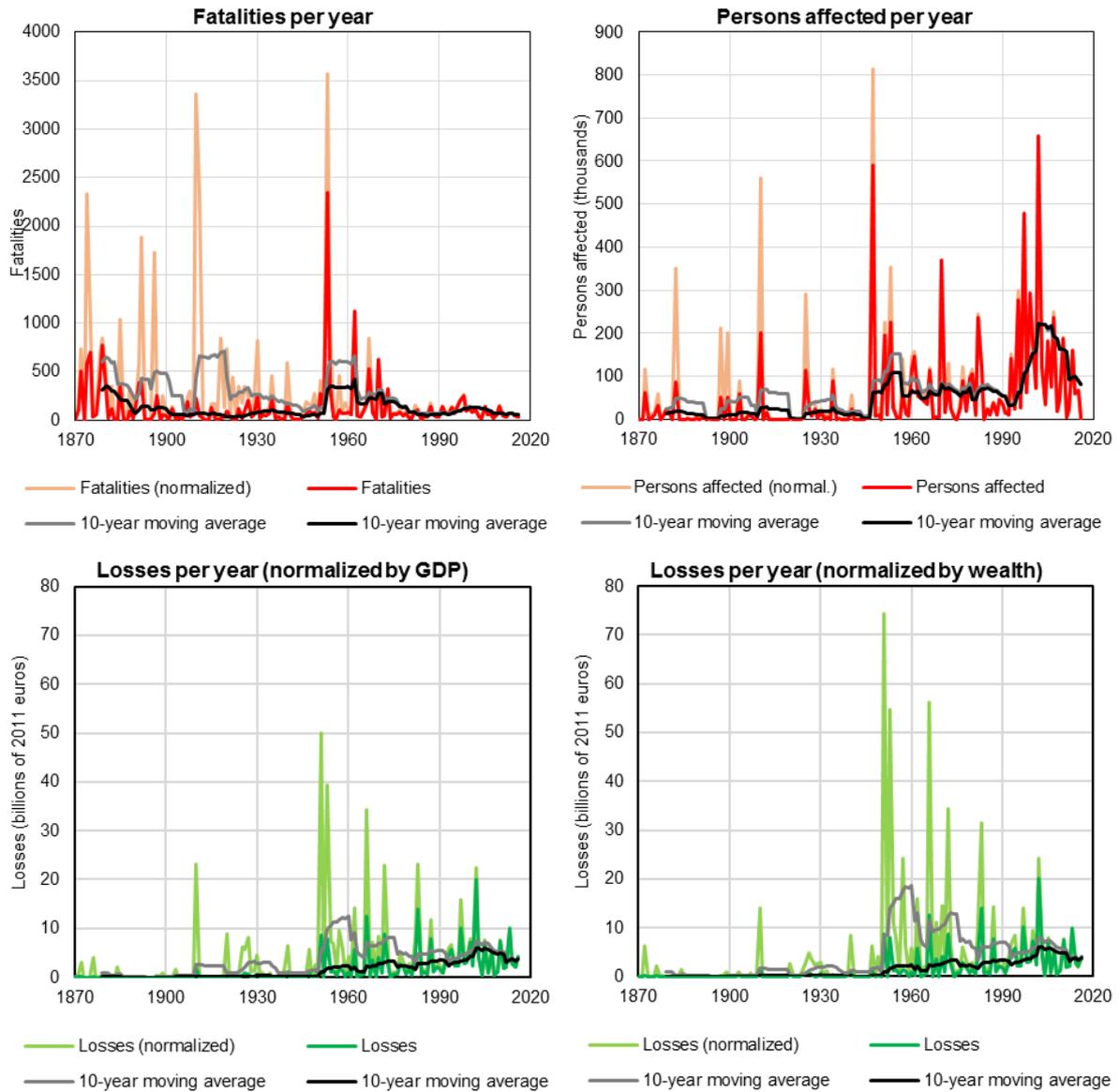

**Figure 3.** Annual number of flood events and their consequences: unadjusted, reported values (dark colours) and normalized values, i.e. adjusted to 2011 levels of exposure (lighter colours).

**Table 1.** Trends in reported, normalized and gap-filled annual losses, during five periods in the historical record. Values are in % per year and equal the rate parameter in Poisson regression. For uncertainty ranges, see Supplementary Fig. 5 and 6.

| Starting year | Reported | | | | | Normalized | | | | Normalized and gap-filled | | | | |
|---|---|---|---|---|---|---|---|---|---|---|---|---|---|---|
| | Events | Area | Fatalities | Affected | Losses | Fatalities | Affected | Losses 1 | Losses 2 | Area | Fatalities | Affected | Losses 1 | Losses 2 |
| 1870 | *1.5 | *1.4 | -0.3 | *2.0 | *3.0 | *-1.1 | *1.1 | *1.5 | *1.4 | *1.6 | *-1.2 | *0.7 | 0.2 | -0.1 |
| 1900 | *1.5 | *2.0 | 0.2 | *2.0 | *2.8 | *-1.4 | *1.2 | 1.0 | 0.9 | *1.8 | *-1.3 | 0.6 | 0.2 | 0.3 |
| 1930 | *1.3 | 1.6 | -0.9 | *1.7 | *2.4 | *-1.8 | 1.1 | -0.1 | 0.3 | *1.7 | -1.8 | 0.4 | -0.5 | -0.0 |
| 1950 | *1.0 | 0.6 | *-3.3 | 1.4 | 1.3 | *-4.6 | 0.8 | *-2.6 | -1.8 | *1.3 | *-4.7 | -0.1 | *-2.3 | *-1.5 |
| 1970 | *1.4 | -1.5 | -1.7 | 1.2 | 1.3 | -1.9 | 0.9 | -1.6 | -0.6 | 1.0 | *-2.0 | 0.3 | -1.2 | -0.3 |

Note: [1] normalized by wealth, [2] normalized by GDP, * significant at α = 0.05.

**Trends in flood losses corrected for missing damage information.** Historical records of flood events often do not contain all or even most of the statistics on flood consequences. Hence, in order to better assess trends in flood losses, gaps in the database were filled using estimates based on an analysis of the dependence structure between all pairs of variables using copulas (see Methods). Gap-filled annual losses are presented in Fig. 4. The difference between the unadjusted and gap-filled data is clearly visible in the graphs; only in case of the number of fatalities are the differences small. This is because there were few gaps in the historical record for the number of fatalities.

The addition of modelled data to the historical record affected many of the observed trends, both compared to reported and normalized losses (Table 1). The trend in inundated area for 1950–2016 becomes statistically significant after gap-filling (1.3% per year), while an opposite trend is indicated for 1970–2016: an annual increase of 1% (not significant) instead of an annual decrease of 1.5%. However, for the entire period 1870-2016, there is little difference in the observed upward trend after gap-filling (1.6% instead of 1.4%). In terms of the number of deaths, there is almost no change in trends, as fatalities decline across the board, with the trend for 1950–2016 reaching -4.7% per year. The number of persons affected before correcting for missing records shows an 0.8-1.2% increase across all considered time periods, while after correction, the trend decreases to at most 0.7%, annually, with a small decline during the period between 1950 and 2016. Only the 1870–2016 trend is statistically significant. Moreover, the normalized monetary value of losses after gap-filling no longer shows a significant trend for the whole period, and losses normalized by wealth increase by only 0.2% per year, while normalized by GDP decline by 0.1% per year. For all other time slices, the general trends are the same as before correcting for missing data.

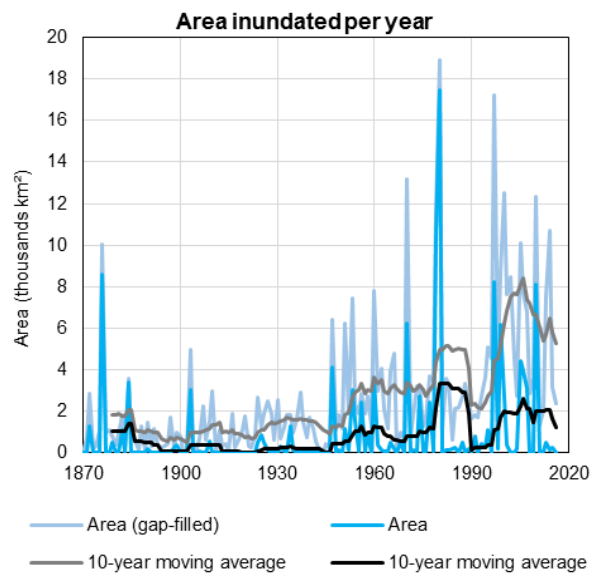

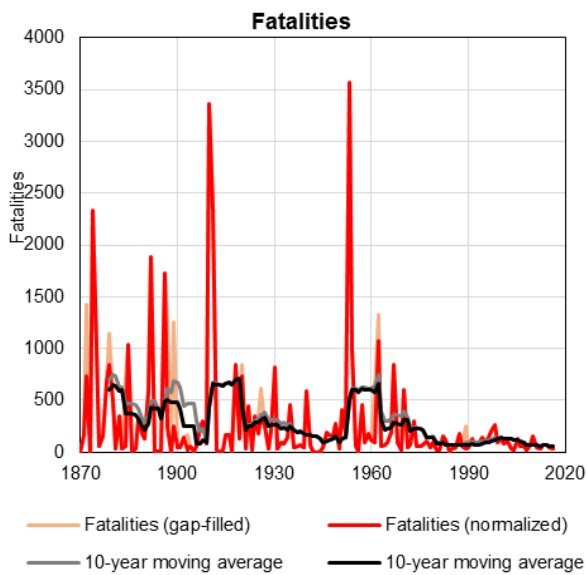

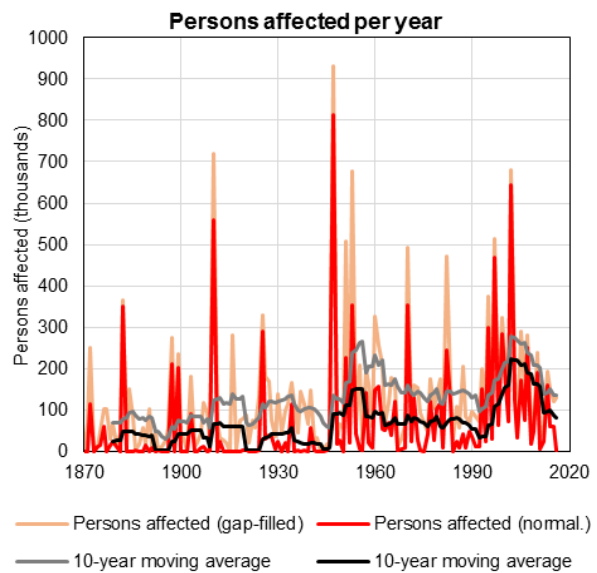

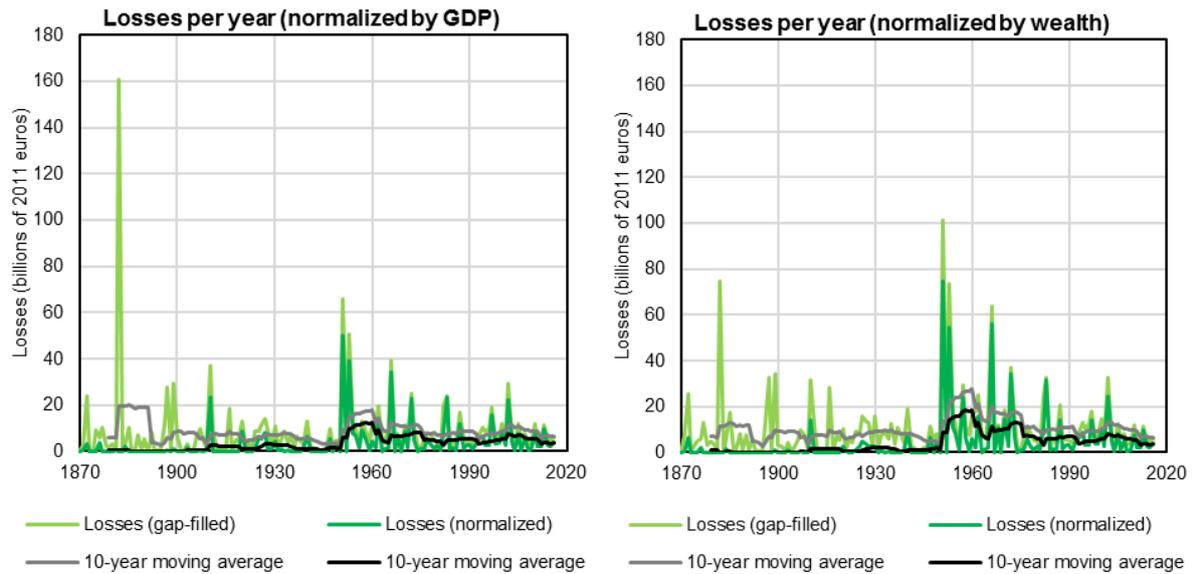

**Figure 4.** Annual consequences of floods with (lighter colours) and without gap-filling (dark colours). All data are normalized except for area inundated, for which normalization was not applied.

**Estimation of underreporting of flood events.** The findings presented here include several uncertainties. One is the completeness of the database of historical floods. In principle, per each extreme flood event in the record, there should also be multiple smaller ones. However, there are relatively few small events recorded in HANZE before about 1950. If we divide the flood events by severity into quintiles (Fig. 5), the smaller the flood, the steeper the observed trend in number of flood events. For example, the annual increase in number of flood events in the uppermost quintile (i.e. largest floods) is 0.3% per year compared to 2% per year for those in the lowest quintile. This finding is also the same when splitting flood events by decile (with less than 0.1% increase per year in the upper 10%). This points to substantial underreporting of smaller floods historically; they are simply not mentioned in contemporary publications referring to historical events. Yet, small floods remain important since they can have a large contribution to overall damages over longer periods of time[53]. In the present, better availability of news reports and government data improves coverage considerably.

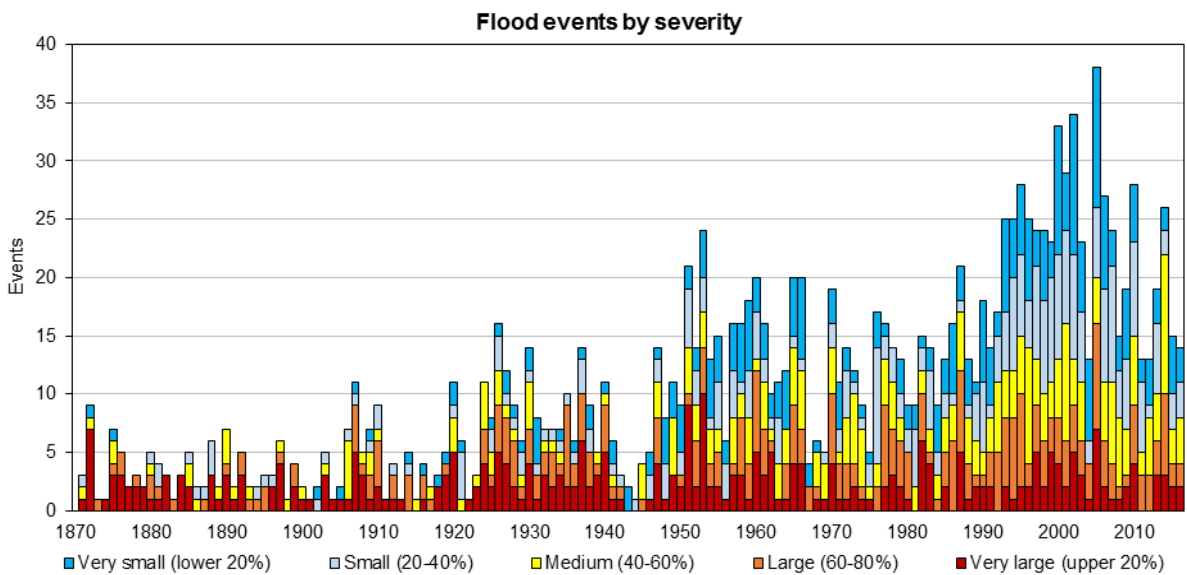

**Figure 5.** Annual number of flood events classified by severity into quintiles. Classification is based on normalized and gap-filled values of losses.

To estimate the quantity of missing information, or underreporting, we adjust the number of events (except those in the upper 20%) before 1990 so that the ratio between number of events in each quintile is the same as after 1990 (see Methods for details). A summary of all adjustments to reported data is presented in Fig. 6. We find that correcting for underreporting diminishes most of the upward trend in number of flood events, whereas it only slightly reduces the growth in area inundated. Yet, given the very small number of recorded flood extents, there is considerable uncertainty in both gap-filling and underreporting correction. The decline in number of fatalities becomes more pronounced with every adjustment and the gap-filled data suggest that number of people affected peaked in the mid-20$^{th}$ century, with no significant trend thereafter. With all corrections applied, a downward trend in financial losses becomes apparent, although for losses normalized by wealth a mid-century peak is indicated. In total, we estimate that flooding affected 0.03% of European population per year on average between 1870 and 2016, and generated losses equal 0.08–0.09% of GDP (depending on normalization variant).

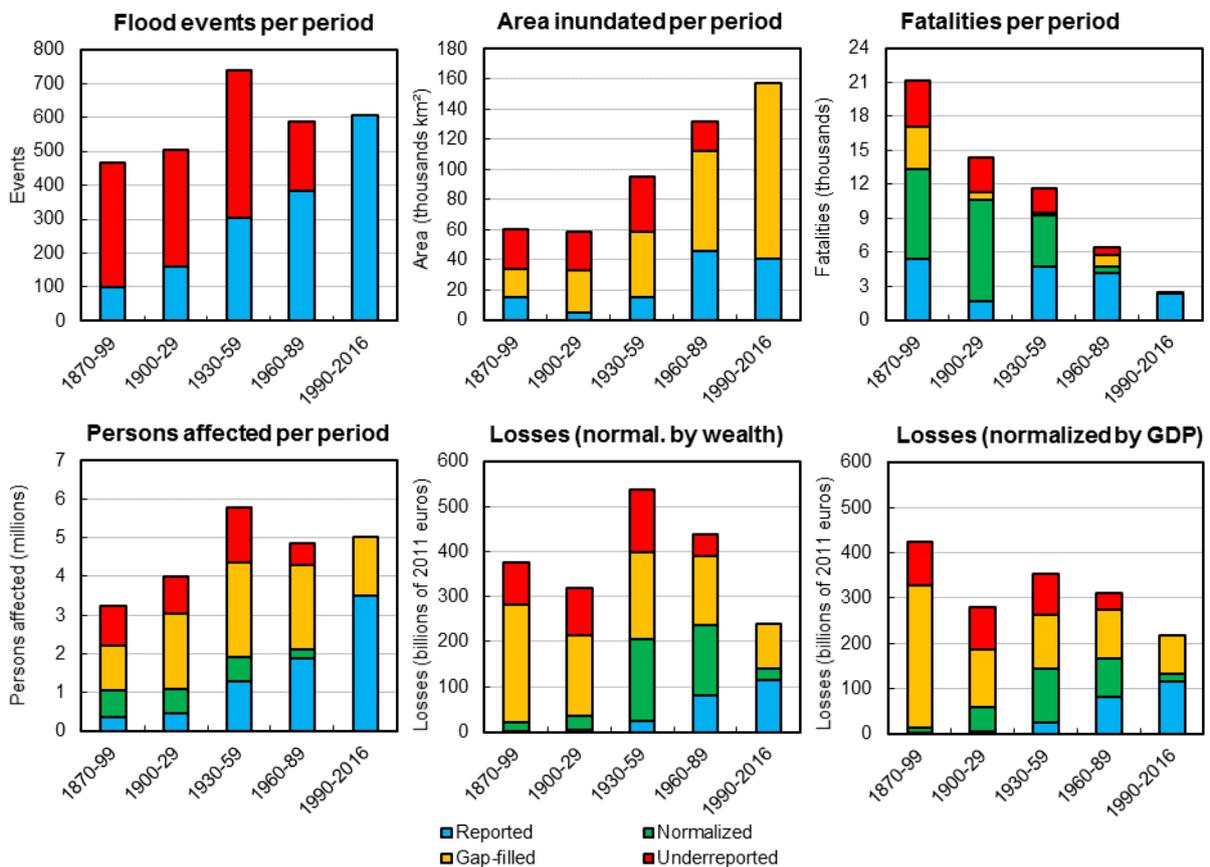

**Figure 6.** Reported number of flood events and their consequences, summed per 30-year periods, with three types of adjustments: normalization, gap-filling of missing loss data (normalized) and estimation of underreporting of small flood events and normalized damages they caused.

**Validation of flood footprints.** Another source of uncertainty is the delineation of flood 'footprints'. Here, we used 100-year flood zones from pan-European modelling carried out in project RAIN, which correspond to the climate and physical geography of the 1971–2000 period. However, we acknowledge that not every flood is a 100-year event, and that the 100-year floodplain boundaries do not remain stationary over time, given, for example, changes in climate, river geometry, urban development, or construction of hydraulic structures. But, detailed, local flood hazard maps and recorded outlines for historical floods are not readily available for all locations in Europe, requiring a proxy for floodplain extent. To validate the assumption that the 100-year is a usable proxy, we recalculated the results for England using flood extents from a comprehensive study by the Environment Agency (EA)[54]. Trends in exposure inside and outside the flood zones are very similar for both pan-European maps from RAIN project and more detailed maps from EA (Fig. 7). The normalized number of affected persons within actual flood outlines recorded by EA yields an annual downward trend for 1946–2016 of 3.5%, compared to a 2% decline using the HANZE flood footprints and reported number of persons affected. However, the records are dominated by just a few events, especially the 1947 Thames valley flood and 2007 country-wide summer flood, hence there is large uncertainty in this comparison. The total (normalized) number of people within EA flood outlines for 1946–2016 is 1.11 million, compared to normalized reported number of people affected in HANZE of 1.19 million.

We also analysed trends in reported annual losses for Poland between 1947 and 2006 based on national government statistics (Supplementary Fig. 8). For inflation-adjusted, but not normalized, losses an annual upward trend of 3.9% per year was found compared to a 4.2% increase in HANZE. Correcting for national GDP growth, reported annual losses still increase by 1.9%. In contrast, normalized and gap-filled data for Poland in HANZE indicate a 2.8% increase per year.

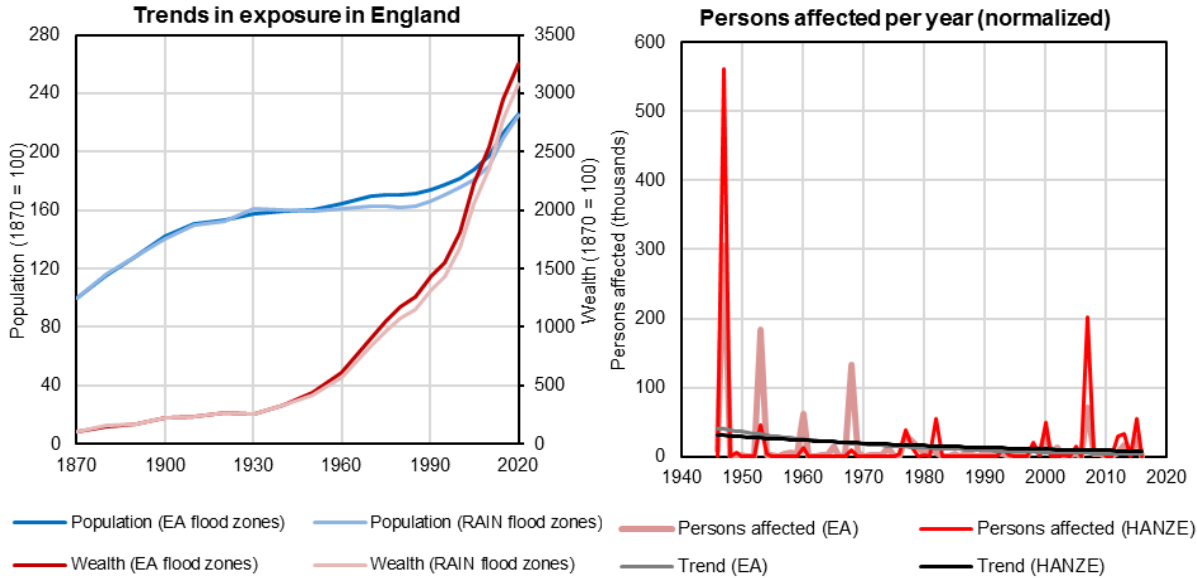

**Figure 7.** Left: trends in population and fixed assets living within 100-year flood zone in England, using Environment Agency (EA) flood risk map and RAIN project map used in this study. Right: estimated persons affected (normalized) in England, compiled by intersecting EA historical flood outlines with HANZE-Exposure population grid, and compared with normalized reported persons affected from HANZE-Events. The trends were calculated using Poisson regression.

## Discussion

This study contains further sources of uncertainty which are less easily quantifiable. For instance, we assume that the flood zones are constant over time. Climate change notwithstanding, many developments may alter local flood hazards, such as river regulation or construction of defences, bypass channels and reservoirs. In case of the latter, we include the erection of large reservoirs in land use, but do not consider their effects on the size of flood zones. Other uncertainties are related to the normalization and gap-filling of damage statistics, though we include the probable margins of error in statistical significance testing (see Methods). Naturally, reported data could also contain many inaccuracies and inconsistencies. For example, there are many variations in the way that the number of people affected are reported across different sources, ranging from the number of evacuees to the number persons whose houses were either inundated or destroyed. Often, only the number of houses affected (flooded, damaged or destroyed) was provided for a given event. In this case, we assumed 4 persons per household, as some other national/international databases also used this assumption. In other cases, there might also be incomplete coverage of financial loss data, in the sense they do not always include all categories of assets. Information on area inundated more often than not refers only to agricultural land flooded rather than complete extent of events.

Nevertheless, the findings presented here are consistent with previous studies. No significant trend was reported for financial losses normalized at country-level for major European floods (1970–2006)[55], major European windstorms (1970–2008)[56], or Spanish floods (insured losses, 1971–2008)[57]. For those time periods, insignificant downward trends were observed in the HANZE gap-filled financial losses normalized by wealth (-0.4 to -0.7% per year). In the United States, an insignificant annual decline of 0.49% was found in flood losses normalized by change in tangible wealth (1932–1997)[58]. This is similar to a 0.12% decline recorded in HANZE during those years for Europe. In Australia, no trend was found in insured losses from weather-related hazards for years 1967–2006, when the losses were corrected for increase in dwelling value[59]; however, in HANZE an insignificant upward trend of 0.2% per year was observed.

Given the one-and-half century timespan of the study, an important question is raised as to whether the results indicate an influence of climate change. In the aforementioned study for the US, trends in precipitation were found to be similar to trends in flood losses per capita. For Europe, we used the 20th Century Reanalysis[60] to obtain trends in the number of extreme precipitation events (return period above 5 years) with a duration from 1 to 7 days. An annual increase of 0.7–1.2% was observed in the data for 1870–2014 (0.8–1.4% when using a 10-year return period). This is slightly below a 1.4% increase both in the (unadjusted) number of flood events and (gap-filled) area inundated. However, when considering underreporting, the number of events and flooded area had likely less pronounced trends. This might indicate an increase in flood hazard caused by climate change and, as a result, a decrease in vulnerability of population and assets. On the other hand, given the significant deficiencies of data on flooded area, uncertainty in the underreporting of smaller flood events and potential bias in reanalysis data, this correlation could be coincidental. The average for Europe also masks large spatial diversity of meteorological and hydrological trends[61], let alone differences in adaption to flood risk.

In future studies, more research could be done on influence of social, political and technical factors on changes in flood vulnerability and risk. In this study, the most significant trend observed was a decline in flood-related fatalities of 1.4% per year since 1870 and 4.3% since 1950. Many technological factors could explain this decrease, such as vast improvements in communication and transportation, which allowed more effective evacuation, rescue and relief operations, and the establishment of meteorological and hydrological agencies, which allowed for continuous observation and forecasting of rainfall and river discharges, improved early warning and disaster preparedness. Moreover, flood prevention, emergency management and disaster relief have largely become permanent government services, in contrast to ad-hoc local arrangements of the past. Dwellings have also become sturdier as brick and concrete is more often used as construction material than timber or adobe. These changes would mostly effect on the number of casualties, but have a relatively limited effect on inundated area, persons affected or economic losses. No trends or slight increase in those variables suggest that there has been no radical improvement in flood prevention since 1870. Still, more data collection is needed, especially to gain more confidence in local hydrologic trends. Only when the climate signal is removed from the data, can the trend in flood vulnerability be computed and the effectiveness of adaptation assessed.

## Methods

**General information**. The HANZE database, used as the basis for this study, includes records of damaging historical floods and a dataset of gridded land cover/use, population, GDP and wealth that allows us to calculate changes in exposure within any given flood 'footprint'. HANZE covers 37 countries and territories in Europe: all 28 European Union member states, all four European Free Trade Agreement members (Iceland, Liechtenstein, Norway and Switzerland), four microstates located in Western Europe (Andorra, Monaco, San Marino and the Vatican) and the Isle of Man. The domain excludes the Canary Islands, Ceuta, Melilla, the Azores, Madeira and Northern Cyprus. Below, a summary of the methodology is presented. Further information about HANZE can be found in the database documentation[62].

**Modelling changes in exposure.** The general concept of the methodology is based on the HYDE database[63][64]. First, two detailed maps of population and land use are compiled for one point in time – 'baseline maps'. Other time points in the past and in the future were calculated based on those baseline maps. Here, the maps refer to the year 2011/12, and have a spatial resolution of 100 m. For the years between 1870–2020 only the total population and land use at NUTS 3 regional level (1353 units)[65] is known. Hence, for each time step, the population and the different land use classes was redistributed inside each NUTS 3 region in order to match the regional totals.

Baseline land cover/use was taken from Corine Land Cover (CLC) 2012, version 18.5a[66] and population from GEOSTAT grid containing figures from 2011 population censuses[67]. The population grid was further refined to 100 m resolution using two disaggregation methodologies described by Batista e Silva et al.[68]. First, the 1 km population was redistributed into land use classes within each grid cell using an iterative 'limiting variable' method (M1 in the aforementioned paper) and CLC 2012 map. Then, population from land use classes was further distributed into 100 m cells proportional to soil sealing (method M3 in Batista e Silva et al.) taken from the Imperviousness 2012 dataset[69].

A database of statistics covering years 1870–2020 at NUTS 3 level was compiled from multiple sources covering population number, percent living in urban areas, persons per households, percent of land covered by croplands and pasture, and area covered by transportation infrastructure. The land use and population distribution was modified starting from the baseline map as follows. Population per urban grid cell was modified according to changes in mean number of persons per household. Surplus population and urban fabric was removed starting with grid cells furthest away from urban centres until it matched regional total of urban population. Then, area covered by industry was changed proportionately to industrial production per capita in constant prices. 'Industrial' grid cells located furthest from the urban centres were removed first when going back in time. Reservoirs were removed completely using the information on year of construction from GRanD database[70]. Area covered by transport infrastructure in a region was changed as defined in the database of historical statistics. 'Infrastructure' grid cells located furthest from the urban centres were removed first when going back in time. Airports were removed completely using the information on year of construction. Airports were identified in CLC 2012 using mostly OurAirports database[71] and year of construction was found in Internet resources. All construction was removed from the land use map for years 1870–2005, otherwise as in the baseline map. The area covered by croplands in a region was adjusted to match the value in the historical statistics, so that the grid cells least suitable for agriculture were removed first, while unutilized grid cells with the highest suitability were added first. Suitability is proportional to slope (from EU-DEM[72]) and crop suitability index for high-input cereals (from FAO[73] Same-ranked grid cells) was disambiguated with distance from urban centres (see text for details). Pastures were redistributed as croplands, but with crop suitability index for high-input alfalfa used instead of cereals. All burnt areas were removed from the land use map for years 1870–2000, otherwise as in the baseline map. If, after application of previous steps some land became unoccupied, it was assumed that this land was covered by the same natural land cover that typical in its nearest neighbourhood, unless no natural land cover could be found, in which case the grid cells were assumed to be covered by forest. Finally, the population of grid cells were changed from urban to non-urban was modified to a land cover-specific value, and the non-urban population was modified according changes in mean number of persons per household. If needed, rural population was added/removed based on distance from urban centres to match historical statistics for a region. The remaining CLC 2012 classes (ports, dump sites, natural water bodies and courses, glaciers etc.) were assumed constant.

As a last step, GDP (compiled at NUTS 3 level with sectoral breakdown) and wealth (non-financial, produced, tangible fixed assets compiled as % of national GDP with sectoral breakdown and then multiplied by GDP at NUTS 3 level) were disaggregated to a 100 m grid. Half of the GDP generated by agriculture (excluding forestry), as well as half of wealth generated by agriculture was distributed proportionally among the population living in agricultural areas. The other half was distributed equally among CLC classes corresponding to agricultural areas. GDP and wealth in forestry were distributed the same way, but using forests. Half of GDP and wealth in industry and services was distributed proportionally to the population in all grid cells, while the other half was distributed equally among specific land use classes where given production is concentrated. The whole wealth in dwellings was distributed proportionally to the population in all grid cells. The entire value of infrastructure, on the other hand, was distributed equally over selected land use classes: urban fabric, airports, ports, roads and railway sites.

**Compiling a database of flood events.** HANZE includes information on past damaging floods that occurred in the study area between 1870 and 2016. Records of flood events were obtained from a large variety of sources, including international and national databases, scientific publications and news reports. Sources are identified per event in the dataset itself (see 'data availability'). Flood events, in order to be included in the database had to fulfil certain criteria. First, at least one of the four damage statistics (area flooded, fatalities, persons affected and monetary value of losses) had to be available for a given event. However, if no persons were known to have been killed in the flood, at least one of the remaining statistics had to be available. Second, available information for a given event had to be good enough to assign month, year, country, regions affected, type and cause of the flood. Insignificant floods, which affected only a small part of one region and had no fatalities, were not included in the database. Floods that were caused by insufficient drainage in disconnected urban areas, floods caused entirely by dam failure unrelated with a severe meteorological event, and floods caused by geophysical phenomena were also excluded. Events affecting more than one country were split in the database per country.

**Flood footprints and normalization.** The extent of each flood event was obtained by intersecting a map of regions affected by an event with the flood map from the RAIN project, available from 4TU.ResearchData repository for river[74] and coastal[75] floods. The flood maps are for a 100-year return period and historical scenario (1971–2000). The floodplain includes all river sections with a catchment area above 100 km$^2$. The map does not include flood defences, therefore constitutes all potentially inundated areas. It should be noted that seven events were not included in the normalization and further analysis due to lack of flood extent data: four flash floods in Malta (where river were too small for inclusion in RAIN flood map) and three coastal floods in Sicily (where no flood risk was indicated in RAIN map).

Normalization was carried out by multiplying reported losses by the relative change in population, GDP or wealth within each event's footprint. As an example we can consider the 1953 North Sea flood in the Netherlands, which caused 1835 fatalities and 4.8 bln euro damages in 2011 prices. Given that the population within the flood's footprint increased by 60% and wealth by 636% between 1953 and 2011, the normalized fatalities will amount to 2930 persons and financial losses to 35.5 bln euro. It is therefore assumed that the vulnerability is constant within the timeframe of the study and all losses would have changed proportionally to local demographic and economic growth.

**Correcting for gaps in historical data availability.** Missing information on losses for events recorded in HANZE database was filled based on correlation between the four variables describing flood damages. Normalized values relative to potential damages within a given flood footprint were used. The empirical distribution of each variable was converted to ranks and the joint distribution of each pair of variables was fitted to five types of copulas (Gaussian, Gumbel, Clayton, Frank and Plackett)[76]. The best-fitting copula for each case was chosen according to the "Blanket Test" described by Genest et al.[77], which uses the Cramèr–von Mises statistic. For a given event and missing data, the available variable that was most highly correlated with the missing particular sample of the variable of interest was used. The conditional copula was sampled 10,000 times to generate samples of the conditional distribution of interest and mean of the conditional damage was used as the estimate of the missing values. The relative damage was the multiplied by total exposure within a given flood event's footprint. The graphs of dependency structures (transformed to standard normal space) are shown in Supplementary Fig. 4 with correlations and best-fitting copula types are included in Supplementary Table 1.

Underreporting of smaller flood events in the past was estimated by transforming normalized and gap-filled damage statistics (with financial losses normalized by wealth only) to ranks (highest to lowest) and dividing the events into quintiles based on their average rank. It was then assumed that the catalogue of events in the upper quintile (20%), i.e., the most severe events, is complete over the entire dataset. For other four quintiles, the catalogue is assumed complete only during the most recent period: 1990–2016. During this period, the ratio of events between four lower quintiles to the highest one was 1.60, 2.02, 2.42 and 2.29 (higher quintile to lower). For other 30-year time periods (1870–1899, 1900–29, 1930–59, 1960–89) the ratio is lower, which was considered to be a function of underreporting of less severe floods (Supplementary Fig. 5). Hence, reported flood events were multiplied by factors necessary to achieve the same ratios between quintiles as in 1990–2016, where the highest quintile was not adjusted as we assume the records of most severe floods are complete. The same factors were applied to multiply flood consequences for all variables.

**Analysing trends in flood risk.** Trends were analysed using Poisson regression, which is better suited for count data than linear regression[78][79]. Statistical significance of the trends presented in the paper was analysed by Monte Carlo simulation. The trend calculated for a given variable (rate parameter of Poisson regression) was compared with 10,000 samples of randomised data series. Those randomised series were annual number of flood events or their consequences, where each flood event had a randomly assigned year from a uniformly distributed interval [1870, 2016]. For each of the 10,000 randomised series the Poisson regression was calculated in order to obtain confidence intervals. The trend for a given variable was considered significant if the rate parameter was higher than in 95% of trends of randomised data series. As an additional check, the *t* test was applied to the calculated trends, yielding the same results at α = 0.05 significance level.

Reported values of variables were then 'normalized', i.e. for each flood footprint the reported value of losses were multiplied by the change in exposure between year of event and 2011 baseline. To test statistical significance in the normalized data series, we first estimated the uncertainty distribution of past exposure. It was assumed to be a log-normal distribution fitted to the empirical distribution of change in exposure between given time point and 2011 within all NUTS 3 regions. This log-normal exposure distribution was sampled to obtain a random value of exposure per given flood event. This sampling was repeated 10,000 times for each flood event to generate a set of randomised data series of annual normalized flood losses. This allowed us to compute uncertainty ranges in normalized data series in Supplementary Fig. 5. Then, a randomised data series were further randomised by assigning a year from a uniformly distributed interval [1870, 2016] to each flood event, as in previous paragraph. The trend was considered significant if it was higher than 95% of randomly-generated trends.

For gap-filled data series, the uncertainty in the modelled data was further incorporated into significance testing. For each missing value of flood loss for a given event, 10,000 samples of marginal distribution of that variable obtained during the copula analysis. This allowed us to compute uncertainty ranges in normalized data series in Supplementary Fig. 6. Like for normalization, the data series incorporating uncertainty of gap-filling were further randomised by assigning a year from a uniformly distributed interval [1870, 2016] to each flood event, as in previous paragraph. The trend was considered significant if it was higher than 95% of randomly-generated trends.

**Validation.** Comparison of exposure and flood losses trend was carried out using two Environment Agency (EA) maps. 'Risk of Flooding from Rivers and Sea', April 2017 version, contains highly-detailed flood zones at several probabilities of occurrence[80]. 'Recorded Flood Outlines', May 2017 version, contains actual flood extents continuously recorded since 1946, with a limited number of events from earlier years as well[81]. The potential flood zones were intersected with population and wealth maps for 1870–2020, and the recorded outlines since 1946 were intersected with the disaggregated baseline population map. Additionally, we compared trends reported annual losses for Poland for 1947–2006 with the trends based on HANZE-Events. Annual losses from Polish sources[82][83][84] were normalized using national GDP series.

Precipitation trends were computed using NOAA-CIRES 20th Century Reanalysis, version 2c[85]. It is a global climate reanalysis for 1851–2014 with a 3-hourly temporal resolution and 2° spatial resolution. A total of 329 grid cells intersect with the study area, for which daily precipitation amounts were extracted for years 1870–2014. For every grid cell a 5-year empirical return period of 1-, 2-, 3-, 5- and 7-day precipitation was calculated and then the number of events which exceeded this threshold was obtained. Finally, this number of extreme events was summed for all grid cells per each year.

**Data availability.** The HANZE database used in this study is publicly available from 4TU.ResearchData with the identifier '10.4121/collection:HANZE' (ref. 46) and from the corresponding author upon reasonable request.

## Acknowledgements
This work received support of project "Bridging the Gap for Innovations in Disaster Resilience" (BRIGAID), which received funding from the European Union's Horizon 2020 research and innovation programme under grant agreement no. 700699. Further support was provided by project "Risk Analysis of Infrastructure Networks in response to extreme weather" (RAIN), which received funding from the European Union's Seventh Framework Programme for research, technological development and demonstration under grant agreement no. 608166.

## Author contributions
D.P. conceived and designed the study, prepared and analysed the data, and wrote the manuscript. A.S. assisted in the interpretation of the data, participated in technical discussions, and helped compose the manuscript. O.M.N. and S.N.J. helped guide the research through technical discussions. All authors revised the manuscript and gave final approval for publication.

# Supplemental Information

**List of figures**

**List of tables**

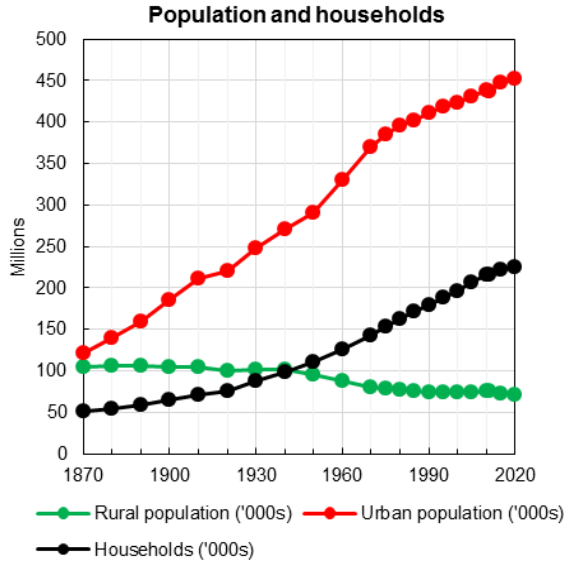
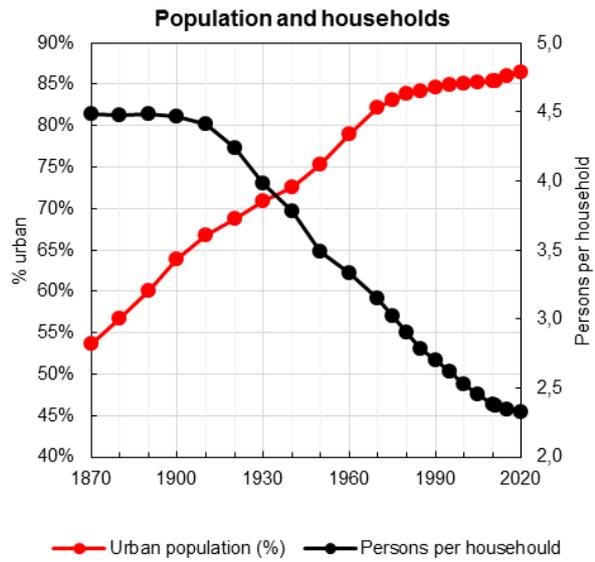
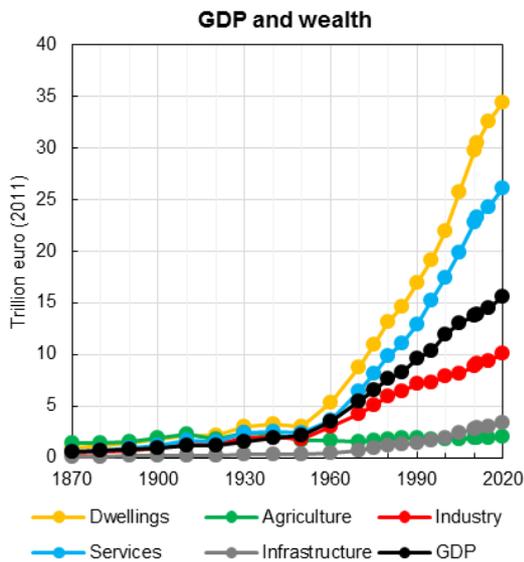
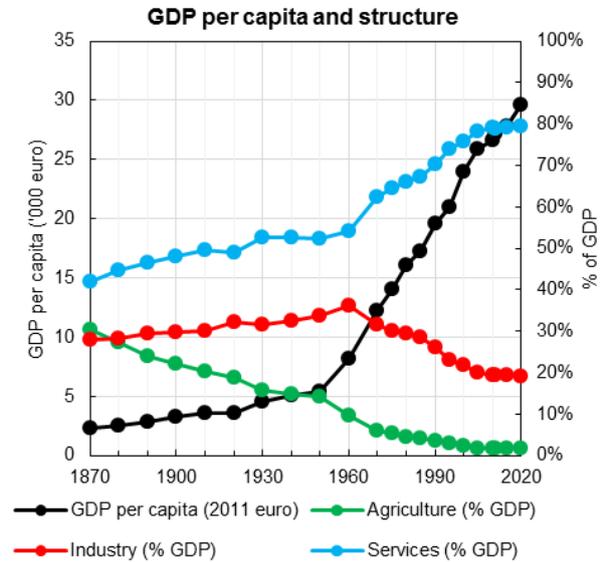
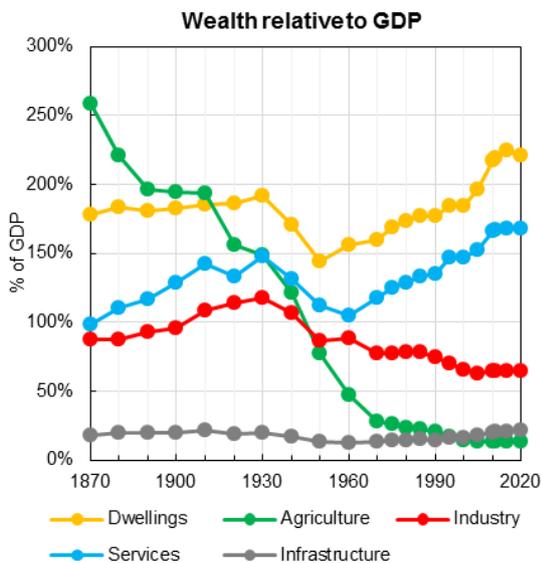
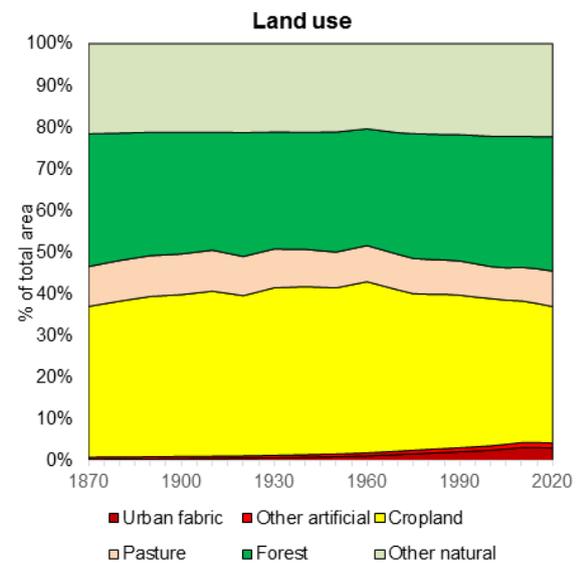

Supplementary Figure 8. Temporal trends for selected socio-economic variables since 1870, including short-term projection up to 2020. Aggregate for 37 European countries and territories. Source: HANZE database.

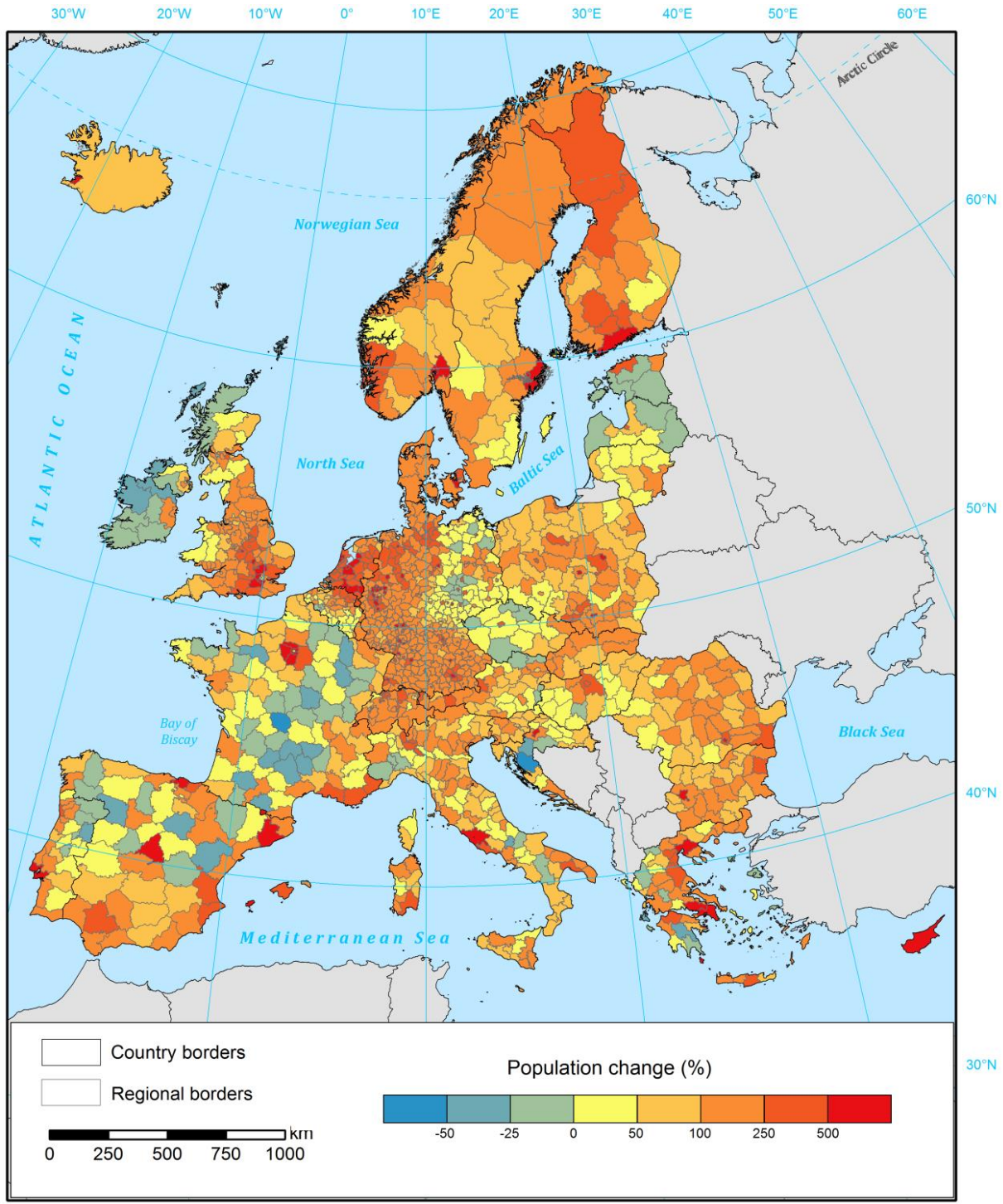

Supplementary Figure 9. Population change (%) by NUTS3 regions from 1870 to 2015. Source: HANZE database.

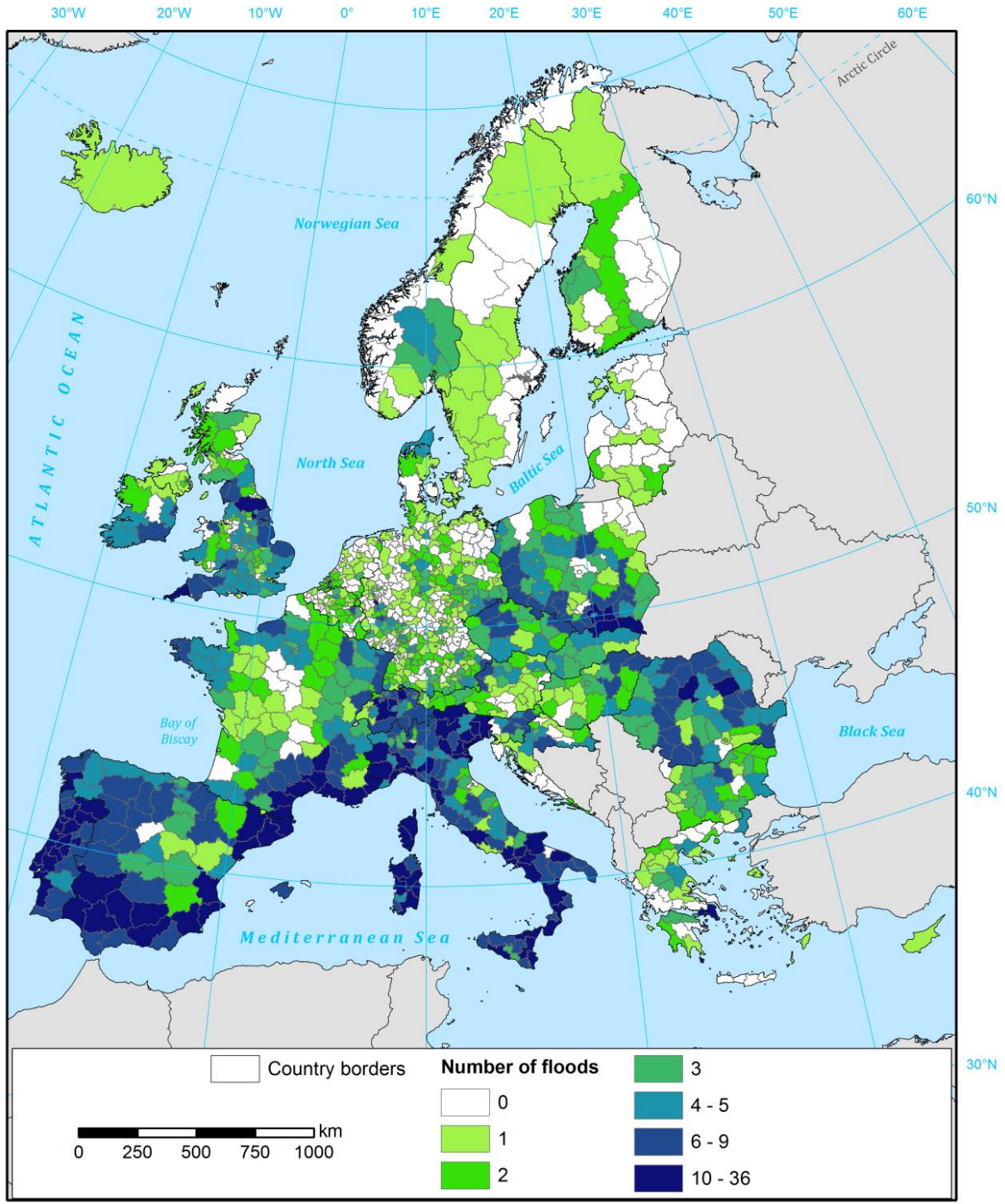

**Supplementary Figure 10. Total number of floods events recorded in HANZE database by NUTS3 regions, 1870–2016.**

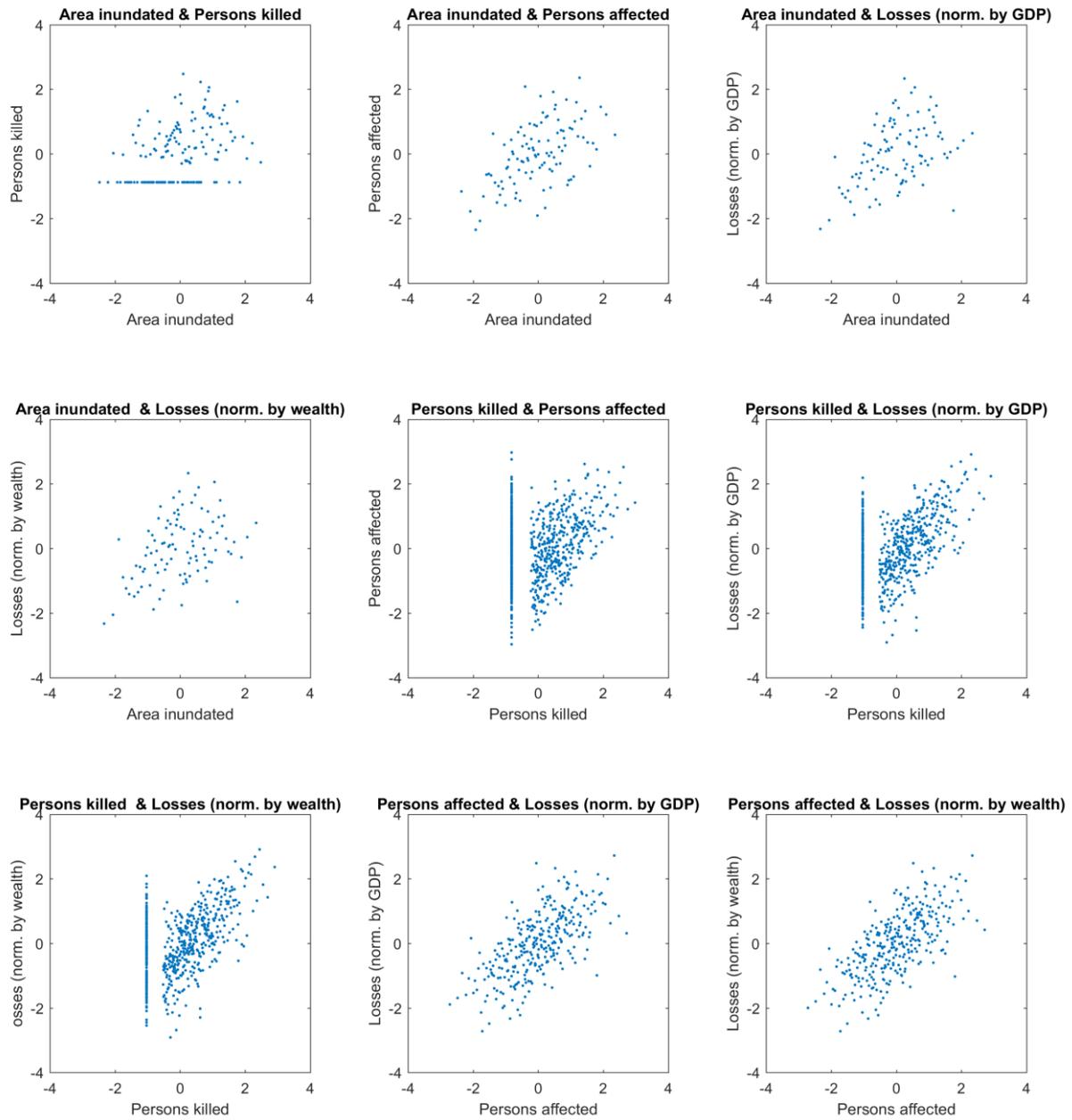

**Supplementary Figure 11.** Dependency between pairs of variables (normalized damage statistics relative to potential exposure per flood footprint) transformed to standard normal.

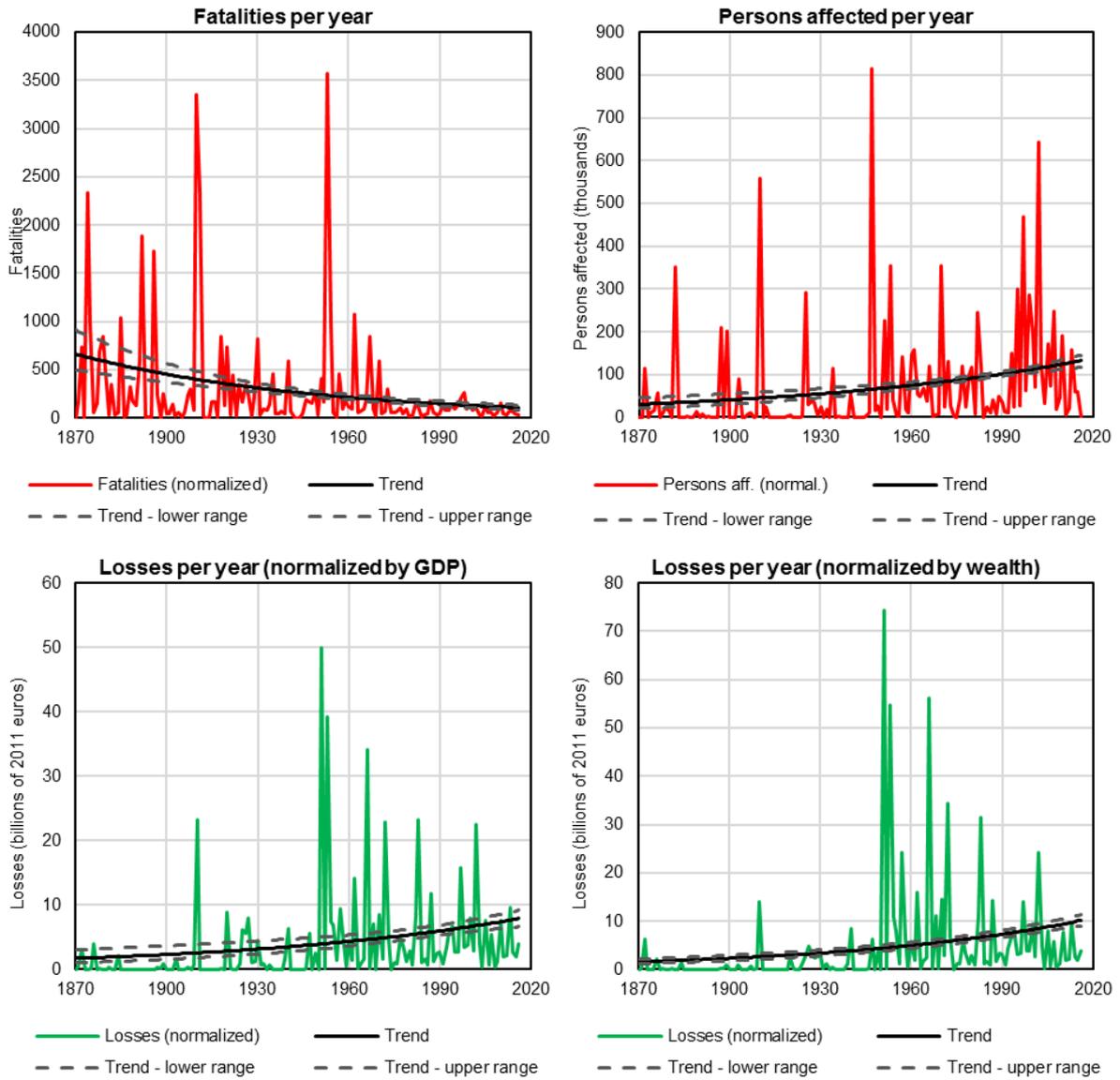

**Supplementary Figure 12. Trends in normalized flood losses with 95% confidence intervals.**

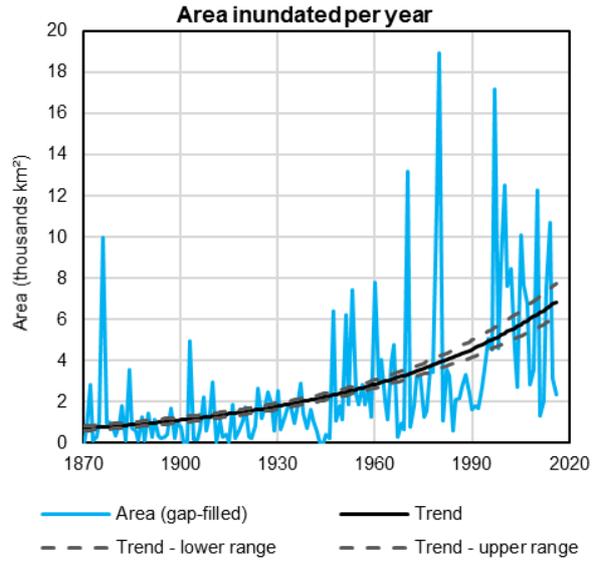

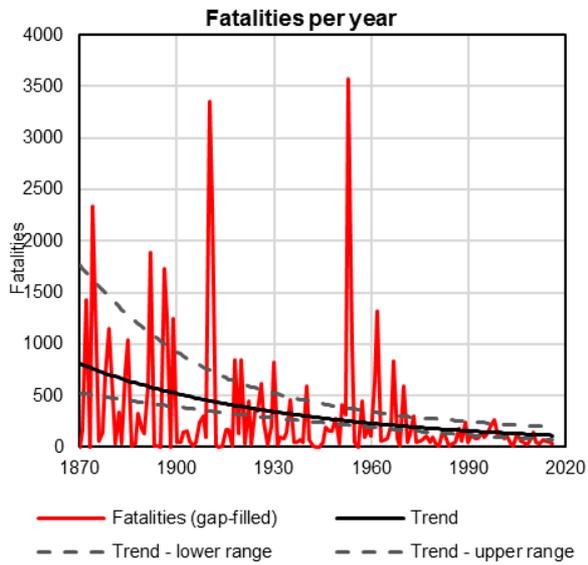
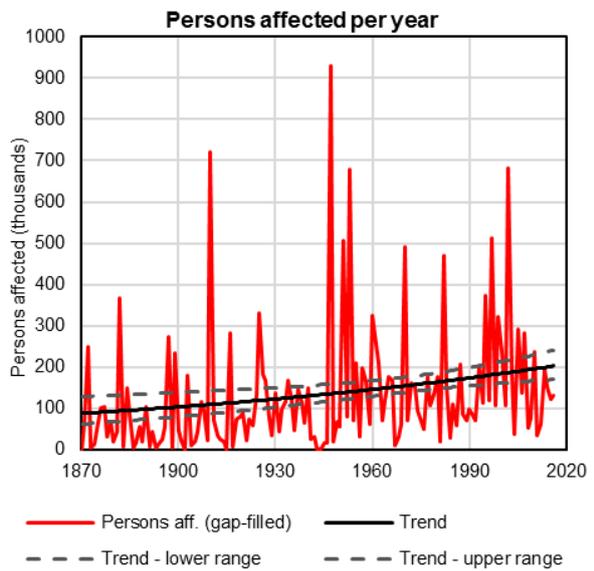

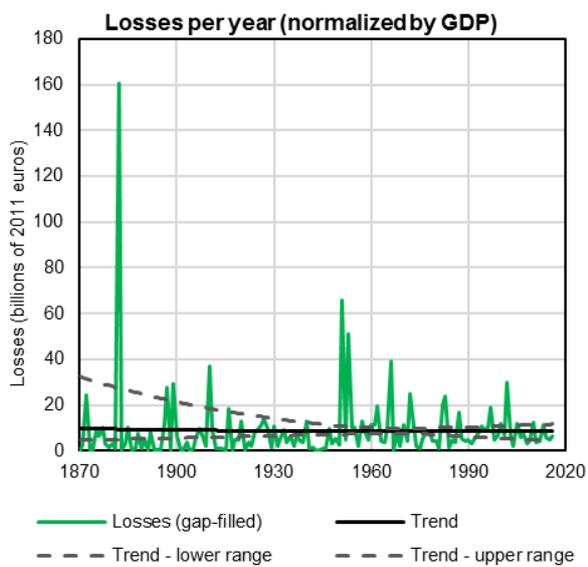
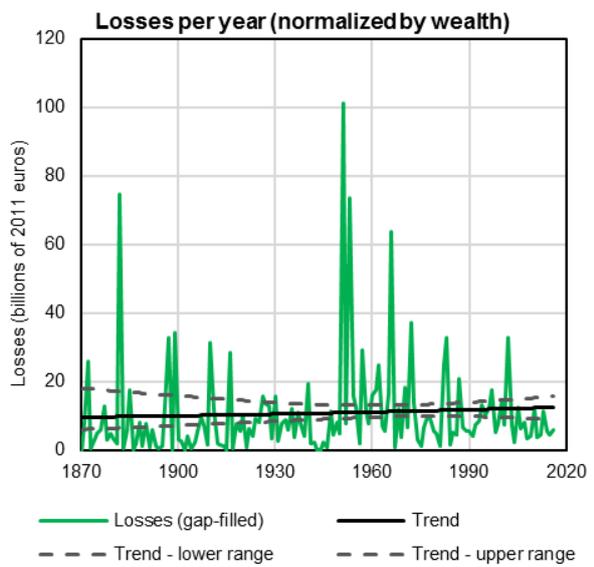

**Supplementary Figure 13. Trends in normalized and gap-filled flood losses with 95% confidence intervals.**

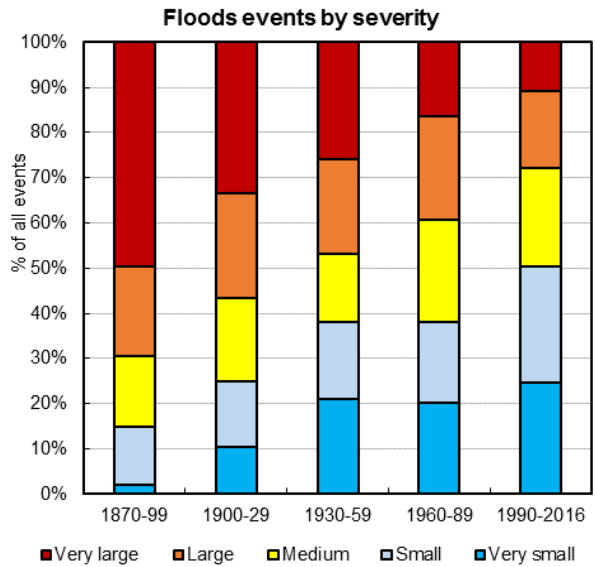

**Supplementary Figure 14. Flood events classified by severity per time period.**

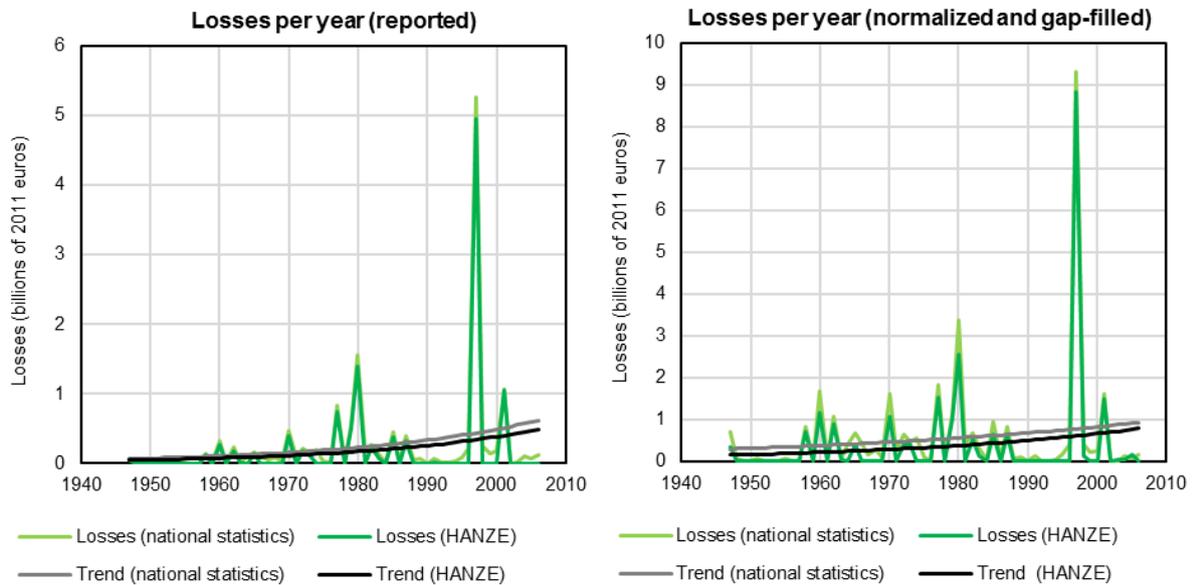

**Supplementary Figure 15. Annual financial losses to floods in Poland, 1947–2006, according to national statistics and HANZE database. The trends were calculated using Poisson regression.**

**Supplementary Table 2. Correlation and best-fitting copulas for pairs of variables (normalized damage statistics).**

| Pair of variables | Spearman's rank correlation | Best-fitting copula type |
|---|---|---|
| Area inundated & Fatalities | 0.352 | Frank |
| Area inundated & Persons affected | 0.527 | Clayton |
| Area inundated & Losses (norm. by GDP) | 0.431 | Clayton |
| Area inundated & Losses (norm. by wealth) | 0.376 | Clayton |
| Fatalities & Persons affected | 0.272 | Normal |
| Fatalities & Losses (norm. by GDP) | 0.469 | Gumbel |
| Fatalities & Losses (norm. by wealth) | 0.473 | Gumbel |
| Persons affected & Losses (norm. by GDP) | 0.667 | Frank |
| Persons affected & Losses (norm. by wealth) | 0.677 | Frank |